\documentclass[aps,prb,nofootinbib,superscriptaddress,floatfix,10pt,twocolumn]{revtex4-2}

\usepackage{graphicx}
\usepackage{amsmath}
\usepackage{bm}
\usepackage{color}
\usepackage{overpic}
\usepackage{changes}
\usepackage[breaklinks=true,colorlinks,citecolor=blue,linkcolor=blue,urlcolor=blue]{hyperref}

\usepackage{siunitx}
\newcommand{\dif}{\mathrm{d}}
\newcommand{\I}{\mathrm{i}}
\let\Re\relax \let\Im\relax
\DeclareMathOperator{\Re}{Re}
\DeclareMathOperator{\Im}{Im}

\begin{document}

\title{Tuning Fermi liquids with polaritonic cavities}
\author{Riccardo Riolo}
\affiliation{Dipartimento di Fisica dell'Universit\`a di Pisa, Largo Bruno Pontecorvo 3, I-56127 Pisa,~Italy}
\author{Andrea Tomadin}
\affiliation{Dipartimento di Fisica dell'Universit\`a di Pisa, Largo Bruno Pontecorvo 3, I-56127 Pisa,~Italy}
\author{Giacomo Mazza}
\affiliation{Dipartimento di Fisica dell'Universit\`a di Pisa, Largo Bruno Pontecorvo 3, I-56127 Pisa,~Italy}
\author{Reza Asgari}
\affiliation{Department of Physics, Zhejiang Normal University, Jinhua, Zhejiang 321004,~China}
\affiliation{School of Quantum Physics and Matter, Institute for Research in Fundamental Sciences (IPM), Tehran 19395-5531,~Iran}
\author{Allan H. MacDonald}
\affiliation{Department of Physics, The University of Texas at Austin, Austin, TX 78712,~USA}
\author{Marco Polini}
\affiliation{Dipartimento di Fisica dell'Universit\`a di Pisa, Largo Bruno Pontecorvo 3, I-56127 Pisa,~Italy}
\affiliation{ICFO-Institut de Ci\`{e}ncies Fot\`{o}niques, The Barcelona Institute of Science and Technology, Av. Carl Friedrich Gauss 3, 08860 Castelldefels (Barcelona),~Spain}
\begin{abstract}
The question of whether or not passive sub-wavelength cavities can alter
the properties of quantum materials is currently attracting a great deal
of attention. In this Article we show that the Fermi liquid parameters
of a two-dimensional metal are modified by cavity polariton modes, and
that these changes can be monitored by measuring a paradigmatic
magneto-transport phenomenon, Shubnikov--de Haas oscillations in a weak
perpendicular magnetic field.  This effect is intrinsic, and totally
unrelated to disorder. As an illustrative example, we carry out explicit
calculations of the quasiparticle velocity
of graphene in a planar van der Waals cavity formed by
natural hyperbolic crystals and metal gates.
The largest effects of the cavity occur when the phonon
polariton modes of the former match energetically the graphene plasmon.
For typical graphene carrier densities this occurs in the Terahertz
spectral range.
\end{abstract}

\maketitle

\paragraph*{Introduction.} Magnetic oscillation phenomena~\cite{Shoenberg,Landau09} like the de Haas--van Alphen effect, oscillations of the magnetization of a metal, 
are a powerful probe of the physics of itinerant electron systems. 
The frequencies of de Haas--van Alphen oscillations, for example, are an historically important 
probe of the Fermi surfaces of metals~\cite{Shoenberg}.
In metallic two-dimensional electron systems (2DESs) 
strong magnetic oscillations appear in the longitudinal resistivity $\rho_{xx}$~\cite{Ando_RMP_1982}
(Shubnikov--de Haas (SdH) oscillations) 
at magnetic fields weaker than those at which the quantum Hall effect occurs. 
Importantly, these can be used to access the Landau Fermi liquid parameters~\cite{Rice_AnnPhys_1965,Pines_and_Nozieres,Mahan_2000,Giuliani_and_Vignale} of  2DESs such as the quasiparticle effective mass $m^\ast$ (or, equivalently, quasiparticle velocity $v^\ast_{\rm F}$) and
the interaction enhanced $g$-factor $g^\ast$. Measurements in ultra-clean systems, including 2DESs confined to silicon inversion
layers~\cite{Fang_PhysRev_1968,Smith_PRL_1972,Pudalov_PRL_2002}, AlAs/AlGaAs quantum wells~\cite{Vakili_PRL_2004,Padmanabhan_PRL_2008,Gokmen_PRL_2008,Gokmen_PRB_2009}, GaAs/AlGaAs quantum wells~\cite{Tan_PRL_2005,Tan_PhysicaE_2006,Chiu_PRB_2011}, ${\rm LaAlO}_3/{\rm SrTiO}_3$ interfaces~\cite{Caviglia_PRL_2010,BenShalom_PRL_2010}, SiGe/Si/SiGe quantum wells~\cite{Melnikov_SciRep_2023}, and atomically-thin materials~\cite{Novoselov_Nature_2005,Zhang_Nature_2005,Elias_NatPhys_2011,Zou_PRB_2011,Fallahazad_PRL_2016}, report large deviations of $m^\ast$ and $g^\ast$ from the non-interacting values $m_{\rm e}$ and $g$.

The aim of this Article is to investigate whether SdH oscillations are affected by the fluctuations of an electromagnetic field confined to a small volume of space by a sub-wavelength~\cite{Novotny_2012,Basov_Nanophotonics_2021,Plantey_ACS_2021} optical cavity. 
The possibility of tailoring the ground state and
transport properties of an electron system solely by altering 
the fluctuations of a \textit{passive} cavity (i.e.~a cavity in the absence of optical pumping) is clearly potentially exciting and has recently captured a great deal of attention~\cite{Bloch_Nature_2022,Schlawin_APR_2022,GarciaVidal_Science_2021,Genet_PT_2021,Rubio_NatureMater_2021}.

In 2DESs, transport is a particularly convenient probe to monitor 
passive cavity tuning of many-electron properties. The pioneering
theoretical work on cavity-induced modifications of magneto-transport
has been performed by Ciuti and co-workers~\cite{Bartolo_PRB_2018,Ciuti_PRB_2021,Arwas_PRB_2023}
and Rubio and co-workers~\cite{Rokaj_2023}. Recent
experiments~\cite{Paravicini_Bagliani_NaturePhys_2019,scalari_science_2012,muravev_prb_2013,maissen_prb_2014,smolka_science_2014,Keller_nanolett_2017,Appugliese_science_2022} have shown that the magneto-transport properties of a 2DES in a 
GaAs/AlGaAs quantum well can be modified by coupling the electronic 
degrees of freedom to sub-wavelength cavities in the Terahertz (THz) spectral range. In the case of split-ring THz resonators, for example, Ref.~\cite{Paravicini_Bagliani_NaturePhys_2019} reports 
a suppression of the amplitude of SdH traces, but no changes in the frequency $B_{\rm F}$,
in the passive regime.\footnote{%
The experimental findings of Ref.~\cite{Paravicini_Bagliani_NaturePhys_2019} have been interpreted by using the theory of Ref.~\cite{Bartolo_PRB_2018}. The latter deals with the static magneto-resistivity tensor of a 2DES coupled to a single cavity mode. In this theory, the authors take into account~\cite{Bartolo_PRB_2018} the coupling between the cyclotron resonance of the 2DES and the cavity mode but neglect the temperature dependence of the amplitude of the SdH oscillations. We believe that any comparison between theory and experiment should take into account finite temperature effects, which strongly modify the amplitude of the SdH oscillations.
}

\begin{figure}[t]
\centering
\includegraphics{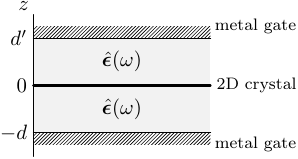}
\caption{A side view of the planar polaritonic cavity analyzed in this work. 
A 2D crystal which hosts a Fermi liquid (thick line) of areal density $n_0$ located at $z=0$ is encapsulated between two hyperbolic dielectrics (light grey) described by frequency-dependent uniaxial permittivity tensors $\hat{\bm \epsilon}(\omega) = {\rm diag}(\epsilon_{x}(\omega),\epsilon_{x}(\omega),\epsilon_{z}(\omega))$. The bottom (top) slab has thickness $d$ ($d^{\prime}$). Metal gates (hatched grey) fill the two half-spaces $z>d^\prime$ and $z < -d$. We assume that the dielectrics are identical.}\label{fig:device}
\end{figure}
\begin{figure}[t]
\centering
\begin{overpic}[percent]{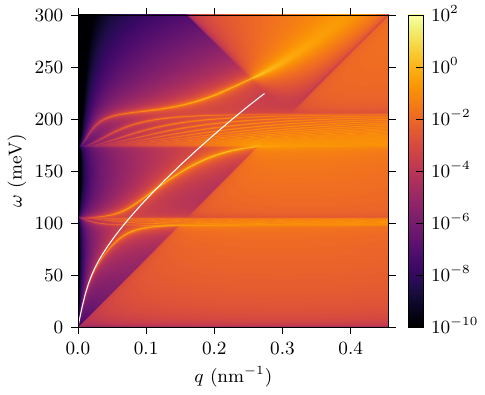}
\put(0,80){\makebox(0,0)[tl]{(a)}}
\end{overpic}
\begin{overpic}[percent]{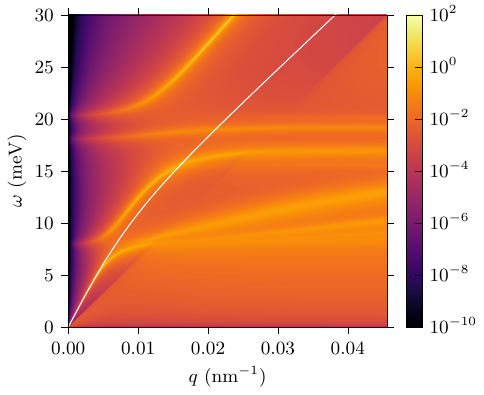}
\put(0,80){\makebox(0,0)[tl]{(b)}}
\end{overpic}
\caption{Dynamical structure factor $S(q,\omega)$ (in units of $k_{\text{F}}/v_{\text{F}}$)
as a function of the wave vector $q$ and frequency $\omega$. Results for graphene encapsulated by 
hBN---Panel (a)---and $\text{Bi}_2\text{Se}_3$---Panel (b).
In both panels, $n_0=\SI{3e12}{cm^{-2}}$,
and $d=d'=\SI{50}{nm}$ (parameters described in Fig.~\ref{fig:device}). The white solid line is the 2DES plasmon 
frequency calculated by artificially substituting
$F(q,0)$ for $F(q,\omega)$
in Eq.~\ref{eq:effective_interaction_uniaxial}, thereby neglecting
cavity dynamics. Notice that the plasmon dispersion is linear at small $q$ because of the presence of metal gates (see Fig.~\ref{fig:device}).}\label{fig:structure}
\end{figure}

Here we focus on the temperature ($T$) dependence of the SdH oscillations.
According to the Lifshitz--Kosevich (LK)
formula~\cite{Lifshitz_1955,Isihara_JPhysC_1986,Sharapov_PRB_2004,Kopelevich_PRL_2004,Kuppersbusch_PRB_2017},
the oscillatory component of the magneto-resistance is given
by\footnote{%
In writing Eq.~(\ref{eq:LK_formula}), we have limited ourselves to the
lowest harmonic.}
\begin{equation}\label{eq:LK_formula}
\frac{\delta \rho_{xx}(B)}{\rho_0} = R(B,T)\cos\left(2\pi\left(\frac{B_{\rm F}}{B}+ \frac{1}{2} + \gamma\right)\right),
\end{equation}
where $\rho_0$ is the resistance at $B=0$, $B_{\rm F} = \phi_0 n_0/N_{\rm f}$ is the frequency of the oscillations in $1/B$, 
$\phi_0=h c/e$ is the magnetic flux quantum ($c$ is the speed of light in vacuum and $e$ the elementary charge), 
$n_0$ is the carrier density,  $N_{\rm f}$ is a degeneracy factor, 
and $\gamma \in [0,1)$ is a Berry phase~\cite{Mikitik_PRL_1999}. 
The amplitude $R(B,T)$ of the SdH oscillations is proportional to (and fully controlled by) the dimensionless quantity
\begin{equation}\label{eq:amplitude}
A(B,T) = \frac{2\pi^2  k_{\rm B} T/\hbar \omega_{\rm c}}{\sinh(2\pi^2 k_{\rm B} T/\hbar \omega_{\rm c})} e^{- \pi/\omega_{\rm c} \tau}.
\end{equation}
Here, $\tau$ is the momentum relaxation time, $m_{\rm c}$ the cyclotron mass, and $\omega_{\rm c} \equiv eB/m_{\rm c}c$ is the 
cyclotron frequency. 
The LK formula is derived under the assumptions of weak fields, i.e.~$\hbar \omega_{\rm c} \ll \mu$, 
where $\mu$ is the chemical potential. The $\tau$-dependent Dingle factor~\cite{Shoenberg} captures
the reduction of the amplitude of the SdH oscillations due to electron-impurity scattering. Note that inelastic processes (such as electron-electron collisions) affect only $R(B,T)$ and not the SdH oscillation frequency $B_{\rm F}$~\cite{Shoenberg}. Eq.~(\ref{eq:amplitude}) is routinely used to fit the measured $T$-dependence of the amplitude of the SdH oscillations at a fixed value of $B$ to extract an experimental value of $m_{\rm c}$.

\begin{figure}[t]
\centering
\begin{tabular}{c}
\includegraphics{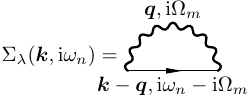}\vspace{0.2 cm}\\
\includegraphics{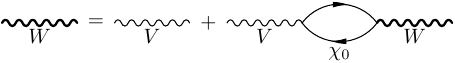}
\end{tabular}
\caption{Feynman diagrams considered in this work.
Top: the irreducible self-energy $\Sigma_\lambda(\bm k,\I\omega_m)$
in the $G0W$ approximation, Eq.~(\ref{eq:sigma_rpa}).
Bottom: the dynamically screened potential $W$ in the random-phase
approximation, as in Eq.~(\ref{eq:dynamical_screening_1}). Here, $V$ represents the Coulomb interaction \textit{dressed by the cavity environment}, and $\chi_0$ the polarization function of the 2DES.
}\label{fig:diagrams}
\end{figure}

We now explain why and how SdH oscillations are modified by a sub-wavelength cavity. 
According to the LK formula~(\ref{eq:LK_formula}), a passive cavity can act via both extrinsic and intrinsic microscopic mechanisms.
The \textit{extrinsic} mechanism is active if the cavity environment alters the momentum relaxation time $\tau$ 
with respect to the case where no cavity is present, and is clearly irrelevant in clean 2DES (i.e.~for $\omega_{\rm c}\tau \gg 1$). The intrinsic mechanism, which is the focus of this Article, is a cavity-induced change in the interaction-induced 
modification of the cyclotron mass $m_{\rm c}$, which we evaluate as $m_{\rm c} = \hbar k_{\rm F}/v^\ast_{\rm F}$, where $k_{\rm F}$ is the Fermi wave number and $v^\ast_{\rm F}$ is the quasiparticle velocity~\cite{Giuliani_and_Vignale}. 
We show by explicit calculation that a sub-wavelength polaritonic cavity---such as
the van der Waals cavity in Fig.~\ref{fig:device}---changes
the quasiparticle weight factor $Z$ and the renormalized
Fermi velocity $v^\ast_{\rm F}$ of the normal Fermi liquid, relative to
their values in vacuum.
The reason is as follows: in the absence of the
cavity, the spectrum or dynamical structure factor $S(q,\omega)$ of a
clean 2DES displays a sharp plasmon mode, which carries a large fraction
of the total f-sum rule spectral
weight~\cite{Pines_and_Nozieres,Giuliani_and_Vignale,Mahan_2000}, and a
continuum of particle-hole excitations. When coupling between the 2DES
and the cavity is allowed, spectral weight flows from the electronic
degrees of freedom to the polariton modes of the cavity,
Fig.~\ref{fig:structure}. This leads ultimately to changes in the Landau
parameters of the 2DES.

Below, we first introduce a general theoretical framework, summarized in Fig.~\ref{fig:diagrams}, that captures this physics
and then present detailed calculations for the case of graphene embedded in the van der Waals cavity sketched in Fig.~\ref{fig:device}
containing metal gates and natural hyperbolic
crystals~\cite{Sun_ACS_2014,Basov_Science_2016,Low_NatMater_2017,Zhang_Nature_2021},
such as hexagonal Boron Nitride
(hBN)~\cite{Dai_science_2014,Caldwell_naturecommun_2014,Li_NatCommun_2015,Dai_NatCommun_2015}
and ${\rm Bi}_2{\rm Se}_3$~\cite{Eva_NatCommun_2021}.\footnote{%
In this work we neglect the role of the surface states of the
topological insulator slabs above and below graphene. Our interest in
${\rm Bi}_2{\rm Se}_3$ here is solely linked to its THz optical
properties.}
A specific illustration of the above mentioned spectral weight transfer is illustrated in Fig.~\ref{fig:structure}, where we plot $S(q,\omega)$~\cite{Pines_and_Nozieres,Giuliani_and_Vignale,Mahan_2000} for graphene in the cavity shown in Fig.~\ref{fig:device}.

\paragraph*{Polariton-mediated effective electron-electron interactions.} We consider an interacting many-electron system 
embedded in a passive cavity, containing (hyperbolic) dielectric media and metal gates.
Because condensed matter is only weakly relativistic, the coupling between matter and electromagnetic degrees of freedom can be separated~\cite{Cohen_1997} into non-radiative and radiative contributions. Non-radiative photons mediate density-density electron-electron interactions, while radiative ones mediate current-current interactions~\cite{andolina_arxiv_2022} that are weaker by $(v_{\rm F}/c)^2$, where $v_{\rm F}$ is the bare Fermi velocity.
For DC transport properties, the effects of photon-mediated density-density interactions 
is dominant with respect to the one of current-current interactions. (Current-current interactions can yield  
large modifications of response functions at \textit{finite} frequency~\cite{Flick_ACS_2019,Amelio_PRB_2021}, 
but have negligible effects on static properties. In the static limit, current-current interactions reduce 
to classical magneto-statics~\cite{nataf_prl_2019,andolina_prb_2020,guerci_prl_2020,mazza_prb_2023}.)

After integrating out cavity degrees of freedom, the effective density-density interaction between electrons
in a sub-wavelength cavity is given by $V({\bm r}, {\bm r}^\prime,
\omega) = e^2 g^{({\rm s})}({\bm r}, {\bm r}^\prime, \omega)$, where
$g^{({\rm s})}({\bm r}, {\bm r}^\prime,
\omega)$~\cite{andolina_arxiv_2022,Asenjo-Garcia_2017}\footnote{%
Green's functions are routinely used~\cite{Landau09} to quantify the
role of vacuum fluctuations as long as one is interested in
electromagnetic fields that vary slowly on the atomic scale. The most
famous example~\cite{Landau09} is probably the Lifshitz theory of the
Casimir-Polder forces.}
is the scalar Green's function for the frequency-dependent Poisson equation~\cite{Landau09}:
\begin{equation}
\label{eq:poisson}
-\partial_\alpha  [\epsilon_{\alpha\beta}(\bm r, \omega) \partial_\beta  \phi(\bm r, \omega)] = 4\pi \rho^{\rm ext} (\bm r, \omega).
\end{equation}
Here, $ \phi(\bm r, \omega)$ is the electrical potential, $\rho^{\rm
ext} (\bm r, \omega)$ is the external charge density, and
$\epsilon_{\alpha\beta}(\bm r, \omega)$ is the frequency-dependent
complex permittivity tensor of the dielectric environment, which is
assumed to be local in space.\footnote{%
Spatial non-locality is important to describe chiral
cavities~\cite{footnote_spatial_nonlocality} and will not be taken into
account in this work. Here, indeed, we neglect the effect of spatial
dispersion since we are interested in the phonon polariton modes of
natural hyperbolic materials, which are very well described by our local
approximation.}
Losses enter the problem via the imaginary part of the permittivity tensor. Metal gates can be included by imposing~\cite{Jackson}  that, at metal-dielectric interfaces, the tangential component of the electric field vanishes and the normal component of the displacement field is $4\pi \sigma$, where $\sigma$ is the surface charge density. Eq.~(\ref{eq:poisson}) can be rigorously derived~\cite{andolina_arxiv_2022} in the generalized Coulomb gauge~\cite{Glauber_PRA_1991}. 

$V({\bm r}, {\bm r}^\prime, \omega)$ describes the interaction between two charges at positions ${\bm r}$ and ${\bm r}^\prime$. In free space, it reduces to the ordinary, instantaneous Coulomb interaction~\cite{Jackson}, i.e.~$v(|{\bm r} - {\bm r}^\prime|)=e^2 / |{\bm r} - {\bm r}^\prime|$. In a sub-wavelength cavity, however, it can be vastly different from $v(|{\bm r} - {\bm r}^\prime|)$. In particular, phonon polaritons in the metallo-dielectric cavity environment yield a retarded frequency-dependent interaction. In the $\omega=0$ limit, $V({\bm r}, {\bm r}^\prime, 0)$ physically describes static screening due to polarization charges in the metallo-dielectric environment. At finite $\omega$, $V({\bm r}, {\bm r}^\prime, \omega)$ encodes the polariton modes of the latter (including losses), which appear as poles of $V({\bm r}, {\bm r}^\prime, z)$ viewed as a function of the complex frequency $z$. 

Given $V({\bm r}, {\bm r}^\prime, \omega)$, one can understand how the properties of the interacting many-electron system change due to the presence of the passive cavity by simply replacing the bare Coulomb interaction $v(|{\bm r}- {\bm r}^\prime|)$ with the retarded interaction $V({\bm r}, {\bm r}^\prime, \omega)$ in the framework of diagrammatic many-body perturbation theory~\cite{Hedin_1969,Mahan_2000,Giuliani_and_Vignale}. This leads to generalized Hedin equations, from which one can calculate, for example, the one-body electron Green's function $G({\bm r}, {\bm r}^\prime, \omega)$, the irreducible self-energy~\cite{Hedin_1969,Mahan_2000,Giuliani_and_Vignale} $\Sigma({\bm r}, {\bm r}^\prime, \omega)$, and the density-density response function~\cite{Mahan_2000,Giuliani_and_Vignale} $\chi_{nn}({\bm r}, {\bm r}^\prime, \omega)$.
Poles in the latter quantity occur at the collective excitation energies of the 2DES/cavity hybrid
system.

\paragraph*{Illustrative example: a planar vdW polaritonic cavity.}
In order to illustrate the power of the approach we just described---while keeping the treatment of the cavity at a fully analytical level---we focus in what follows on a planar sub-wavelength cavity based on a vdW heterostructure~\cite{Geim_Nature_2013}, which is sketched in Fig.~\ref{fig:device}. The active material, where one measures magneto-transport, is a 2D crystalline material located at $z = 0$. This is encapsulated between two homogeneous and uniaxial dielectric slabs of different thicknesses, the slab on top (bottom) having thickness $d^\prime$ ($d$).  The electrical permittivity tensor of these dielectrics is $\hat{\bm \epsilon}(\omega) = {\rm diag}(\epsilon_{x}(\omega),\epsilon_{x}(\omega),\epsilon_{z}(\omega))$. Note, finally, the presence of top and bottom metal gates. Without uniaxial dielectrics we would be dealing with an ordinary Fabry-Pérot cavity~\cite{andolina_arxiv_2022}.

Setups like the one in Fig.~\ref{fig:device} are routinely fabricated in laboratories throughout the world. The most studied case~\cite{Geim_Nature_2013} is the one in which the 2D material is single-layer or Bernal-stacked bilayer graphene and the dielectric material is hBN~\cite{Dai_science_2014,Caldwell_naturecommun_2014,Li_NatCommun_2015,Dai_NatCommun_2015}. Infinite possibilities are clearly possible: many other 2D conducting materials can be chosen (including twisted 2D materials) as well as low-loss dielectrics. For illustration purposes, below we will focus on single-layer graphene (SLG) encapsulated between two natural hyperbolic dielectrics~\cite{Sun_ACS_2014,Basov_Science_2016,Low_NatMater_2017,Zhang_Nature_2021}.

In order to find $V({\bm r}, {\bm r}^\prime, \omega)$ \textit{analytically}, we follow an approach first proposed by Keldysh~\cite{Keldysh_1979} and recently used in Ref.~\cite{Tomadin_prl_2015} in the context of vdW heterostructures. Since the setup in Fig.~\ref{fig:device} is translationally invariant in the $\hat{\bm x}$--$\hat{\bm y}$ plane, the scalar Green's function $g^{({\rm s})}(z,z^\prime, |{\bm r}_{\|} - {\bm r}^\prime_{\|}|, \omega)$ depends only on $|{\bm r}_{\|} - {\bm r}^\prime_{\|}|$. Here, ${\bm r}_{\|}$ and ${\bm r}^\prime_{\|}$ denote 2D positions in the $\hat{\bm x}$--$\hat{\bm y}$ plane. The effective electron-electron interaction (EEEI) that is relevant for the many-body physics of the 2DES located at $z=0$ is $V(0,0,|{\bm r}_{\|} - {\bm r}^\prime_{\|}|, \omega) = e^2 g^{({\rm s})}(0,0, |{\bm r}_{\|} - {\bm r}^\prime_{\|}|, \omega)$.
We are therefore naturally led to introduce its 2D Fourier
transform\footnote{%
The 2D Fourier transform is
$V(q,\omega) \equiv \int\dif^2{\bm r}e^{-\I\bm q\cdot\bm r}
V(0,0,r,\omega)$.}
$V(q,\omega)$ with respect to
$r \equiv |{\bm r}_{\|} - {\bm r}^\prime_{\|}|$.
Once again, the frequency dependence of $V(q,\omega)$ encodes all the information about polaritons while its $\omega \to 0$ limit~\cite{andolina_arxiv_2022} encodes the physics of static screening due to the presence of metal gates and polarization charges.

We find that the polariton-mediated propagator of the planar
vdW cavity sketched in Fig.~\ref{fig:device} is given by $V(q, \omega)
\equiv v_q F(q,\omega)$,
where $v_q \equiv 2\pi e^2/q$ is the non-retarded 2D Coulomb
propagator~\cite{Giuliani_and_Vignale} and 
\begin{equation}\label{eq:effective_interaction_uniaxial}
F(q, \omega) \equiv \frac{2\sinh{\left[ q\sqrt{\frac{\epsilon_x(\omega)}{\epsilon_z(\omega)}}d \right]}\sinh{\left[ q\sqrt{\frac{\epsilon_x(\omega)}{\epsilon_z(\omega)}}d^\prime \right]}}{\epsilon_z(\omega)\sqrt{\frac{\epsilon_x(\omega)}{\epsilon_z(\omega)}}\sinh{\left[ q\sqrt{\frac{\epsilon_x(\omega)}{\epsilon_z(\omega)}}(d + d^\prime) \right]}}
\end{equation}
is a form
factor~\cite{Tomadin_prl_2015,Woessner_naturemater_2015,Lundeberg_Science_2017,Alonso_NatureNano_2017}\footnote{%
Eq.~(\ref{eq:effective_interaction_uniaxial}) has been derived by
treating the top and bottom metal gates as perfect conductors. The
authors of Ref.~\cite{Kim_NatureCommun_2020} transcended the perfect
conductor approximation.
This is relevant in the case of graphite gates or, more in general, when $q_{\rm TF}d$ and $q_{\rm TF}d^\prime$ are small parameters. Here, $q_{\rm TF}$ is the Thomas-Fermi screening wave number~\cite{Giuliani_and_Vignale} of the metal gate, which is proportional to the metal's density of states. The perfect conductor approximation is asymptotically exact in the limit $q_{\rm TF}d, q_{\rm TF}d^\prime \to \infty$.}
due to the presence of the dielectrics and metal
gates. Optical phonons in the dielectrics are responsible for the frequency dependence of $F(q,\omega)$. The components of the uniaxial dielectric tensor are usually parametrized via a Lorentz oscillator model. In the case of a single mode, one has
\begin{eqnarray} \label{eq:BN_dielectric}
\epsilon_\ell(\omega) &=& \epsilon_{\ell}(\infty)
+ \frac{\epsilon_{\ell}(0) - \epsilon_{\ell}(\infty)}{
1 - (\omega/\omega_{\ell}^{\rm T})^2
- \I \gamma_\ell \omega/(\omega_{\ell}^{\rm T})^2},
\end{eqnarray}
with $\ell = x$ or $z$. Here, $\epsilon_{\ell}(0)$ and
$\epsilon_{\ell}(\infty)$ are the static and high-frequency dielectric
constants, respectively, while $\omega^{\rm T}_{\ell}$ is the transverse
optical phonon frequency in the direction $\ell$. The longitudinal
optical phonon frequency $\omega^{\rm L}_{\ell}$ satisfies the
Lyddane--Sachs--Teller relation $\omega^{\rm L}_{\ell} =
\omega_{\ell}^{\rm T}
\sqrt{\epsilon_{\ell}(0)/\epsilon_{\ell}(\infty)}$.
The quantity
$\gamma_\ell$ physically describes losses in the dielectric.
Eq.~(\ref{eq:BN_dielectric}) describes hBN 
accurately and realistic values of the parameters can be found e.g.~in Ref.~\cite{Woessner_naturemater_2015}. 
Multi-mode parametrizations of $\epsilon_\ell(\omega)$ for ${\rm Bi}_2{\rm Se}_3$ can be found in Ref.~\cite{Eva_NatCommun_2021}.

An inspection of Eq.~(\ref{eq:BN_dielectric}) in the limit $\gamma_\ell \to 0$ shows that a dielectric described by this frequency-dependent permittivity tensor is hyperbolic: in the lower (upper) reststrahlen band, which is defined by the inequality $\omega^{\rm T}_{z} < \omega < \omega^{\rm L}_{z}$ ($\omega^{\rm T}_{x} < \omega < \omega^{\rm L}_{x}$), the quantity $\epsilon_x(\omega)/\epsilon_z(\omega)$ takes negative values. Inside the reststrahlen bands, the dressed propagator $V(q, \omega)$ therefore displays poles, which physically correspond to standing hyperbolic phonon polaritons (SHPPs). In the case of negligible losses (i.e.~for $\gamma_\ell=0$), they can be found by looking at the zeroes of the denominator in Eq.~(\ref{eq:effective_interaction_uniaxial}):
\begin{equation}\label{eq:standing_phonon_polaritons}
q_n(\omega)= n \frac{\pi}{(d+d^\prime)\sqrt{|\epsilon_{x}(\omega)/\epsilon_{z}(\omega)|}},
\end{equation}
with $n=1,2, \dots$\footnote{%
The poles of $V(q,\omega)$ are described by Eq.~(\ref{eq:standing_phonon_polaritons}) if and only if $d$ and $d'$ are incommensurate. On the contrary, if $d/d'=m/m'$, the numerator has zeros when
$n$ is an integer multiple of $(m+m')$ and $V(q,\omega)$ has no poles for such values of $n$.
}. Note that the polariton
wavelength
$\lambda_{\rm p}\equiv 2\pi/q_n(\omega) \ll \lambda_0$, where 
$\lambda_{0} = 2\pi (c/\omega)$
is the free-space wavelength. The fact that $\lambda_{\rm p}/\lambda_0\ll 1$ is a distinctive feature of sub-wavelength polaritonic cavities. We now proceed to quantify how the Landau parameters of single-layer graphene (SLG) are modified by these modes.

\paragraph*{Fermi liquid theory in a graphene cQED setup.} The Landau
parameters of a normal Fermi liquid are controlled by the one-body
electron Green's
function~\cite{Hedin_1969,Mahan_2000,Giuliani_and_Vignale}.  In a
(massless Dirac fermion) continuum-model of SLG~\cite{Kotov_RMP_2012},
the physical (i.e.~retarded) electron Green's function $G^{\rm
ret}_{\lambda} ({\bm k}, \omega)$ is a function of a conserved wave
vector ${\bm k}$ and depends on a band index $\lambda=\pm$. It satisfies
the Dyson equation (setting $\hbar =1$), $G^{\rm ret}_\lambda({\bm
k},\omega) = [\omega - \xi_{{\bm k}, \lambda} - \Sigma^{\rm
ret}_\lambda({\bm k},\omega)]^{-1}$, where $\xi_{{\bm
k},\lambda}=\lambda v_{\rm F} k - \mu$ are single-particle band energies
measured from the chemical potential $\mu$ and $\Sigma^{\rm
ret}_\lambda({\bm k}, \omega)$ is the retarded
self-energy~\cite{Mahan_2000,Giuliani_and_Vignale}. The latter quantity
needs to be approximated. In a normal Fermi liquid, a good approximation
is the so-called $G0W$
approximation~\cite{Mahan_2000,Giuliani_and_Vignale,Hedin_1969,Quinn_Ferrell,Hedin_physrev_1965,Hybertsen_PRB_1986}\footnote{%
In the history of the theory of weakly-correlated electron liquids (such
as 3D and 2D parabolic-bands electron gases and graphene), a
plethora of authors has brought up the issue of whether or not one
should calculate the Green's function self-consistently.
An account of these interesting
discussions can be found in Chapter 8 of
Ref.~\cite{Giuliani_and_Vignale}.
The summary of these investigations is that, when vertex corrections are neglected as in the $G_0W$ approximation,
it is much better to \textit{avoid} self-consistency.
For a 3D electron gas see, for example Ref.~\cite{G0W-validity}.
Self-consistent calculations, indeed,
tend to yield an increase in the quasiparticle weight $Z$ and broad
(rather than sharp) plasmon satellites, in contradiction with
experimental data.}
in which the electron self-energy is expanded to \textit{first} order in
the dynamically screened Coulomb interaction $W(q, \I\Omega)$. The
latter is approximated at the level of the random phase approximation
(RPA)~\cite{Mahan_2000,Giuliani_and_Vignale}. The corresponding Feynman
diagrams are displayed in Fig.~\ref{fig:diagrams}. (The same
approximation can be used when translational invariance is not at
play~\cite{Hedin_1969}  and $G$, $\Sigma$, and $W$ depend on ${\bm r}$,
${\bm r}^\prime$, and $\omega$.)

In the case of SLG, the continuum-model $G0W$ self-energy~\cite{Polini_ssc_2007,Polini_prb_2008}
\begin{multline}\label{eq:sigma_rpa}
\Sigma_\lambda({\bm k}, \I\omega_n) = - k_{\rm B} T\sum_{\lambda^\prime}
\int \frac{\dif^2{\bm q}}{(2\pi)^2}\sum_{m=-\infty}^{+\infty}W(q,\I\Omega_m) \\
\times F_{\lambda \lambda^\prime}(\theta_{{\bm k}, {\bm k}-{\bm q}}) G^{(0)}_{\lambda^\prime}({\bm k}-{\bm q}, \I\omega_n - \I\Omega_m),
\end{multline}
where $\omega_n = (2n+1)\pi k_{\rm B} T$ is a fermionic Matsubara frequency, the sum runs over all the bosonic Matsubara frequencies $\Omega_m =2 m \pi k_{\rm B} T$, $F_{\lambda \lambda^\prime}(\theta_{{\bm k}, {\bm k}-{\bm q}}) = [1 + \lambda  \lambda^\prime \cos{(\theta_{{\bm k}, {\bm k}-{\bm q}})}]/2$ is the SLG chirality factor, $\theta_{{\bm k}, {\bm k}-{\bm q}}$ is the angle between ${\bm k}$ and ${\bm k}-{\bm q}$, $G^{(0)}_{\lambda}({\bm k}, \I\omega_n) = 1/(\I\omega_n - \xi_{{\bm k}, \lambda})$ is the non-interacting temperature Green's function, and
\begin{equation}\label{eq:dynamical_screening_1}
W(q,\I\Omega)= \frac{V(q, \I\Omega)}{1- V(q, \I\Omega)\chi_0(q, \I\Omega)}
\end{equation}
is the dynamically screened interaction on the imaginary frequency axis. Here, $\chi_{0}(q, \I\Omega)$ is the SLG non-interacting density-density response function, evaluated on the imaginary-frequency axis~\cite{Barlas_PRL_2007}.
The validity of this non-self-consistent $G0W$ approximation for graphene
has been accurately checked against ARPES experimental
data, see e.g.~Refs.~\cite{Bostwick_2010,Walter_2011}.

We emphasize that the $G0W$ theoretical framework presented in Eqs.~(\ref{eq:sigma_rpa}, \ref{eq:dynamical_screening_1}) is valid also in the presence of losses in the cavity.  In this case, however, care needs to be exercised in defining the quantities that enter Eq.~(\ref{eq:dynamical_screening_1}). Indeed, any thermal Green's function $O(\I\Omega)$ satisfies~\cite{Landau09}
\begin{equation}
O(\I\Omega) = 
\begin{cases}
O^{\rm ret}(\I\Omega), & \text{for $\Omega>0$,} \\
O^{\rm adv}(\I\Omega), & \text{for $\Omega<0$,} \\
\end{cases}
\end{equation}
where $O^{\rm ret}(\omega)$ and $O^{\rm adv}(\omega)$ are the corresponding retarded and advanced Green's functions, respectively. Note that, $O^{\rm adv}(z) = [O^{\rm ret}(z^*)]^*$.
\begin{figure}
\centering
\begin{overpic}[percent]{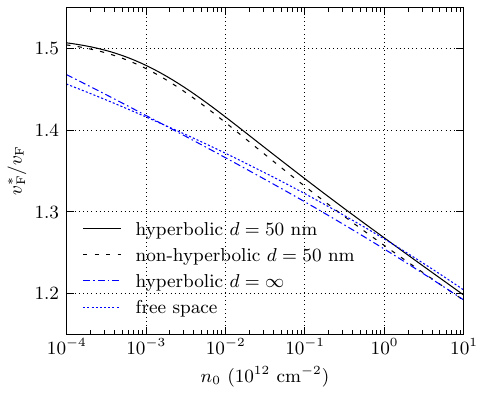}
\put(0,80){\makebox(0,0)[tl]{(a)}}
\end{overpic}
\begin{overpic}[percent]{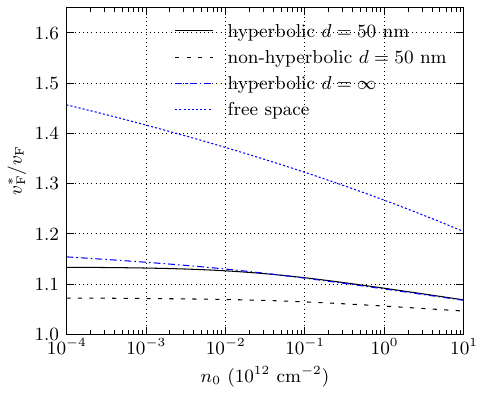}
\put(0,80){\makebox(0,0)[tl]{(b)}}
\end{overpic}
\begin{overpic}[percent]{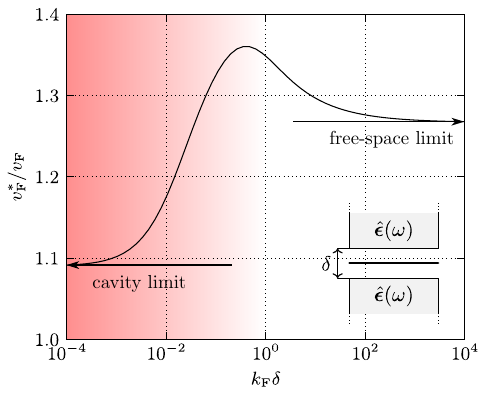}
\put(0,80){\makebox(0,0)[tl]{(c)}}
\end{overpic}
\caption{%
Plot of the quasiparticle Fermi velocity $v^*_{\text{F}}/v_{\text{F}}$
defined in Eqs.~(\ref{eq:v_star_dyson}, \ref{eq:unscreened_velocity}, \ref{eq:screened_velocity}, \ref{eq:full_velocity})
as a function of the electron density $n_0$
for (a) hBN and (b) $\text{Bi}_2\text{Se}_3$.
In both cases, $d^\prime=d$ and the values of $d$ have been indicated in the legend.
Panel (c) shows $v^*_{\text{F}}/v_{\text{F}}$ as a function of the
dimensionless parameter $k_{\rm F} \delta$ where $\delta$ is the distance between two semi-infinite ($d=d'\to\infty$) $\text{Bi}_2\text{Se}_3$ slabs. The inset shows the device configuration. As in Fig.~\ref{fig:device}, the thick black line represents graphene. No metal gates are present in this device. For $k_{\rm F}\delta\gg 1$, we recover the results labelled by ``free space'' in panel (b). For $k_{\rm F}\delta \lesssim 1$ the bulk polariton modes in $\text{Bi}_2\text{Se}_3$ couple strongly to the graphene plasmon and this is therefore the polaritonic ``cavity QED'' regime studied in this work. For $k_{\rm F}\delta\to 0$, we recover the results labelled by ``hyperbolic $d=\infty$'' in panel (b).}\label{fig:vF}
\end{figure}

We also introduce the retarded density-density response function in the RPA approximation
\begin{equation}
\chi_{nn}(q, \omega) = \frac{\chi_{0}(q, \omega)}{1 - V(q,\omega)\chi_0(q,\omega)},
\end{equation}
where $\chi_{0}(q, \omega)$ is the non-interacting (retarded) polarization function on the real-frequency axis~\cite{Shung_PRB_1986,Wunsch_NJP_2016,Hwang_PRB_2006} and the polariton-mediated EEEI $V(q,\omega)$ has been introduced above.  The dynamical structure factor $S(q,\omega)$ plotted in Fig.~\ref{fig:structure} is related to $\chi_{nn}(q, \omega)$ by the fluctuation-dissipation theorem~\cite{Pines_and_Nozieres,Mahan_2000,Giuliani_and_Vignale}, i.e.~$S(q,\omega) = - (\hbar/\pi)\Theta(\omega)\Im \chi_{nn}(q, \omega)$, where $\Theta(\omega)$ is the Heaviside step function.

Note that we can express $W(q,\I\Omega)$ in Eq.~(\ref{eq:dynamical_screening_1}) as:
\begin{equation}\label{eq:dynamical_screening_2}
W(q,\I\Omega)=V(q, \I\Omega) + V^2(q, \I\Omega)\chi_{nn}(q, \I\Omega).
\end{equation}
The first term in the previous equation is responsible for the exchange interaction between a quasiparticle at the Fermi energy and the occupied Fermi sea. The second term, instead, represents the interaction with particle-hole and \textit{collective} virtual fluctuations. Both terms are dressed by SHPPs.

Landau Fermi liquid parameters are controlled by the physical (i.e.~retarded) self-energy $\Sigma^{\rm ret}_+({\bm k},\omega)$, which can be obtained from the analytical continuation $\I \omega_n \to \omega + \I 0^+$ from imaginary to real frequencies. For definiteness, we consider the case of electron-doped SLG with positive chemical potential $\mu$ (results for $\mu<0$ are identical because of particle-hole symmetry). The quasiparticle weight factor $Z$ and the renormalized Fermi velocity $v^\ast_{\rm F}$ (in units of the bare value $v_{\rm F}$) can be expressed in terms of the wave-vector and 
frequency derivatives of the retarded self-energy $\Sigma^{\rm ret}_+({\bm k},\omega)$ evaluated at the Fermi surface ($k=k_{\rm F}$) and at the quasiparticle pole $\omega=\xi_+({\bm k})$~\cite{Mahan_2000,Giuliani_and_Vignale}:
\begin{gather}\label{eq:Z_def}
Z= \frac{1}{1-\left.\partial_{\omega} \Re\Sigma^{\rm ret}_+({\bm k},\omega)\right|_{k=k_{\rm F},\omega=0}}, \\
\label{eq:v_star_dyson} 
\frac{v^\ast_{\rm F}}{v_{\rm F}}
=\frac{\displaystyle 1+(v_{\rm F})^{-1}\left.\partial_k \Re\Sigma^{\rm ret}_+({\bm k},\omega)\right|_{k=k_{\rm F},\omega=0}}{1-\left.
\partial_{\omega} \Re\Sigma^{\rm ret}_+({\bm k},\omega)\right|_{k=k_{\rm F},\omega=0}}.
\end{gather}
Following some standard manipulations~\cite{Giuliani_and_Vignale,Quinn_Ferrell}, the retarded self-energy can be expressed in a form convenient for numerical evaluation, as the sum $\Sigma^{\rm ret}_\lambda({\bm k},\omega) =\Sigma^{\rm res}_\lambda({\bm k}, \omega)+\Sigma^{\rm line}_\lambda({\bm k}, \omega)$ of a contribution from the interaction of quasiparticles at the Fermi energy, the \textit{residue} contribution $\Sigma^{\rm res}_\lambda({\bm k}, \omega)$, and a contribution from interactions with quasiparticles far from the Fermi energy and via both exchange and virtual fluctuations, the \textit{line} contribution $\Sigma^{\rm line}_\lambda({\bm k}, \omega)$. In the zero-temperature limit
\begin{multline}\label{eq:residue_t_0}
\Sigma^{\rm res}_\lambda({\bm k},\omega) =\sum_{\lambda^\prime=\pm}\int \frac{\dif^2 {\bm q}}{(2\pi)^2}
W(q, \omega-\xi_{\lambda^\prime}({\bm k} - {\bm q}))\\
\times F_{\lambda \lambda^\prime}(\theta_{{\bm k}, {\bm k}-{\bm q}})
\left[ \Theta(\xi_{\lambda^\prime}({\bm k} - {\bm q}))
- \Theta(\xi_{\lambda^\prime}({\bm k} - {\bm q}) - \omega) \right]
\end{multline}
and
\begin{multline}
\label{eq:line_t_0_better}
\Sigma^{\rm line}_\lambda({\bm k}, \omega) =-\sum_{\lambda^\prime=\pm}
\int \frac{\dif^2 {\bm q}}{(2\pi)^2}F_{\lambda \lambda^\prime}(\theta_{{\bm k}, {\bm k}-{\bm q}}) \\
\times \int_{-\infty}^{+\infty}\frac{\dif\Omega}{2\pi}\frac{W({\bm q},\I\Omega)}{\omega - \I \Omega -\xi_{\lambda^\prime}({\bm k} - {\bm q})}.
\end{multline}
In what follows we will calculate the following three quantities:
\begin{subequations}
\begin{align}
&\frac{v^\ast_{\text{F}}}{v_{\text{F}}}\biggr\vert_{\text{free\,space}} \equiv \frac{v^\ast_{\text{F}}}{v_{\text{F}}}\biggr\vert_{F(q,\omega)\to 1},\label{eq:unscreened_velocity}\\
&\frac{v^\ast_{\text{F}}}{v_{\text{F}}}\biggr\vert_{\text{non-hyperbolic}} \equiv \frac{v^\ast_{\text{F}}}{v_{\text{F}}}\biggr\vert_{F(q,\omega)\to F(q,0)},\label{eq:screened_velocity}\\
&\frac{v^\ast_{\text{F}}}{v_{\text{F}}}\biggr\vert_{\text{hyperbolic}} \equiv \frac{v^\ast_{\text{F}}}{v_{\text{F}}}\biggr\vert_{F(q,\omega)}.\label{eq:full_velocity}
\end{align}
\end{subequations}
The quantity in Eq.~(\ref{eq:unscreened_velocity}) physically represents the quasiparticle Fermi velocity (measured in units of the bare Fermi velocity $v_{\rm F}$) in graphene, in the total absence of screening stemming from nearby dielectrics and metal gates. It is mathematically calculated by forcing the full dynamical propagator $V(q,\omega)$ to coincide with the instantaneous Coulomb free-space propagator, i.e.~$V(q,\omega) \to v_{q}$.  The quantity $v^\ast_{\rm F}|_{\text{free space}}$ is clearly measurable~\cite{Elias_NatPhys_2011}. 

The quantity in Eq.~(\ref{eq:screened_velocity}) physically represents
the quasiparticle Fermi velocity (in units of $v_{\rm F}$) in a graphene
sheet embedded in the cavity sketched in Fig.~\ref{fig:device}. However,
it is calculated by forcing $V(q,\omega)$ to attain its static value,
i.e.~$V(q,\omega) \to V(q,0)=v_q F(q,0)$. The quantity in Eq.~(\ref{eq:screened_velocity}) therefore includes static screening from the dielectrics and top/bottom metal gates. 
Notice that: i) in the $V(q,\omega) \to V(q,0)$ limit, top and bottom dielectrics are \textit{not} hyperbolic and ii) 
$v^{\ast}_{\rm F}|_{\text{non-hyperbolic}}$ is not measurable but it is an important theoretical construct, as emphasized below. 

Finally, the quantity in Eq.~(\ref{eq:full_velocity}) is the quasiparticle Fermi velocity (in units of $v_{\rm F}$) in a graphene sheet embedded in the cavity sketched in Fig.~\ref{fig:device} and calculated with the full retarded propagator $V(q,\omega)$. This last quantity includes the role of SHPPs.

\begin{figure}[t]
\centering
\includegraphics{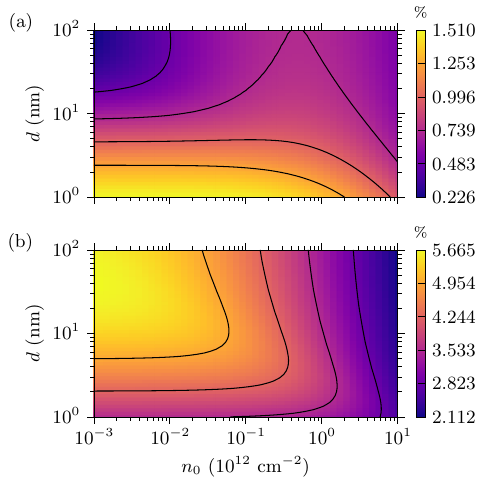}
\caption{Cavity QED effect on the quasiparticle Fermi velocity $v^*_{\rm F}$. Color plots of
$\Delta_{\rm QED}$---as defined in Eq.~(\ref{eq:FOM})---as a function of the electron density $n_0$ and dielectric thickness $d=d'$. Panel (a): the case of hBN. Panel (b): the case of $\text{Bi}_2\text{Se}_3$.
}\label{fig:percentage}
\end{figure}

\paragraph*{Numerical results and discussion.}
We now present our main numerical results. In Fig.~\ref{fig:vF} we show
the dependence of the quasiparticle Fermi velocity $v^\ast_{\rm F}$ (in
units of $v_{\rm F} = 10^6~{\rm m}/{\rm s}$) on carrier density $n_0$,
for $d = d' = 50~{\rm nm}$. Panel (a) in this figure refers to SLG
encapsulated in hBN. In this case and at high electron densities, the
value of $v^*_{\rm F}$ calculated with the retarded EEEI $V(q,\omega)$
differs little from the result $v^*_{\rm F}|_{\text{free space}}$
obtained with the instantaneous Coulomb interaction $v_{q}$.
In the intermediate-to-low density regime (i.e.~for
$10^8~{\rm cm}^{-2} < n_0 < 10^{11}~{\rm cm}^{-2}$), however, the
density dependence of $v^*_{\rm F}|_{\text{hyperbolic}}$ is different
from that of $v^*_{\rm F}|_{\text{free space}}$. While the latter
displays the Gonz{\'a}lez-Guinea-Vozmediano logarithmic
growth~\cite{Kotov_RMP_2012,Elias_NatPhys_2011}, the former saturates to
a constant value. This is a qualitative effect, which may be verified
experimentally, stemming from the different dependence of $V(q,\omega)$
on $q$ with respect to the bare 2D Coulomb
interaction $v_q$. While $v_q$ behaves as $1/q$ at any length scale,
 $V(q,\omega)$ is strongly screened by the metallic gates and
saturates to constant value in the limit $dk_{\text{F}},d'k_{\text{F}}\to0$,
i.e.~$\lim_{dk_{\text{F}},d'k_{\text{F}}\to0}V(q,
\omega)=4\pi e^2 d d^\prime/[\epsilon_{z}(\omega)(d+d^\prime)]$.
Going to densities lower than $10^8~{\rm cm}^{-2}$
(i.e.\ approaching the charge neutrality point $n_0=0$),
the logarithmic growth of $v^*_{\rm F}|_{\text{free
space}}$ overwhelms the saturating behavior of $v^*_{\rm
F}|_{\text{hyperbolic}}$.

In Fig.~\ref{fig:vF} we also present results for the asymptotic regime
$d=d^\prime \to \infty$ (distance between the metal gates is sent to
infinity). The fact that the effect of SHPPs persists also in this limit
indicates that polariton-mediated cavity QED effects do
not require the presence of metallic mirrors
and is peculiar to the case of density-density interactions.
In the case of current-current interactions mediated by radiative
photons~\cite{Ashida_PRL_2023}, fingerprints of SHPPs disappear in the
limit $d=d^\prime \to \infty$ (see Fig.~3 of
Ref.~\cite{Ashida_PRL_2023}).
Since the effect persists in the absence 
of metallic mirrors, the reader may wonder why we talk about ``cavity'' effects.  
The key point is that, in our setup, the dielectric slabs 
precisely act as cavity mirrors that modify the properties 
of the material placed in between them by shaping the 
electromagnetic vacuum in the vicinity of the material. 
To see this, we introduce a new length scale in the 
problem,  i.e.~the distance $\delta$ between the top and bottom 
dielectric slabs in the limit $d=d'\to \infty$,
and place the graphene halfway between the dielectrics\footnote{%
In the case of the device geometry illustrated in Fig.~\ref{fig:vF}(c), the electromagnetic form factor has the following expression:
\begin{equation*}
F(q,\omega)
= \frac{\epsilon_z(\omega)\sqrt{\frac{\epsilon_x(\omega)}{\epsilon_z(\omega)}}
\cosh{\left[q\sqrt{\frac{\epsilon_x(\omega)}{\epsilon_z(\omega)}}\frac{\delta}{2}\right]}
+
\sinh{\left[q\sqrt{\frac{\epsilon_x(\omega)}{\epsilon_z(\omega)}}\frac{\delta}{2}\right]}
}{\epsilon_z(\omega)\sqrt{\frac{\epsilon_x(\omega)}{\epsilon_z(\omega)}}
\sinh{\left[q\sqrt{\frac{\epsilon_x(\omega)}{\epsilon_z(\omega)}}\frac{\delta}{2}\right]}
+
\cosh{\left[q\sqrt{\frac{\epsilon_x(\omega)}{\epsilon_z(\omega)}}\frac{\delta}{2}\right]}
}.
\end{equation*}}
---see the inset in Fig.~\ref{fig:vF}(c).
In the main body of Fig.~\ref{fig:vF}(c) we show the ``cavity'' effect, 
i.e.~the fact that we can tune the quasiparticle velocity $v^\ast_{\rm F}$ by tuning the distance $\delta$ between the two dielectric slabs. Indeed, for $\delta \to \infty$, the polariton modes do not couple to the graphene plasmon and one recovers the ``free space'' limit---see Eq.~(\ref{eq:unscreened_velocity}). Viceversa, tuning $\delta$ towards the $k_{\rm F}\delta \lesssim 1$ regime, boosts the strength of the plasmon-polariton coupling, yielding the strongest renormalization of the Fermi velocity.

An important figure of merit to quantify the role of virtual polaritonic excitations (i.e.~the role of genuine QED effects) on the quasiparticle Fermi velocity is the following:
\begin{equation}\label{eq:FOM}
\Delta_{\rm QED} = \frac{v^*_{\text{F}}\rvert_{\text{hyperbolic}}-v^*_{\text{F}}\rvert_{\text{non-hyperbolic}}}{v^*_{\text{F}}\rvert_{\text{non-hyperbolic}}}~.
\end{equation}
The larger $\Delta_{\rm QED}$ the larger the role of virtual SHPPs. The smaller $\Delta_{\rm QED}$  the larger is the (trivial) role of static screening from polarization charges in the dielectric and metal gates. In other words, large values of $\Delta_{\rm QED}$ certify that changes in the quasiparticle Fermi velocity in a given cavity are not a trivial consequence of static screening. Of course, we have checked that, in the device configuration shown in the inset of Fig.~\ref{fig:vF}(c), $\Delta_{\rm QED} \to 0$ in the limit $k_{\rm F}\delta \gg 1$.

The quantity $\Delta_{\rm QED}$ is plotted in Fig.~\ref{fig:percentage}(a) for the case of SLG encapsulated in hBN. We clearly see that $\Delta_{\rm QED}$ is small, on the order of $1.5\%$ at best, and weakly dependent on $n_0$ and dielectric thickness $d$. We believe that the reason is that hBN reststrahlen bands occur at very high energies, the lower one starting at $\omega =\omega^{\rm T}_{z} \simeq 97~{\rm meV}$. On the contrary, the SLG plasmon carries a large spectral weight (in fact, the total spectral weight) at small values of $q$ and $\omega> v_{\rm F} q$. This ``mismatch'' in the $(q,\omega)$ plane between the SLG plasmon and the hBN (lower) reststrahlen band suppresses spectral flow from the matter degrees of freedom to the phonon polariton ones.

To show that this is indeed the case, we have also calculated the
quasiparticle Fermi velocity $v^*_{\rm F}$ for SLG encapsulated in ${\rm
Bi}_2{\rm Se}_3$. Results for this case are reported in
Fig.~\ref{fig:vF}(b). We immediately see that the cavity plays a much
stronger role, both at the static and polaritonic levels. The fact that
$v^*_{\rm F}|_{\text{non-hyperbolic}}$ is much smaller than $v^*_{\rm
F}|_{\text{free space}}$ stems from the fact that ${\rm Bi}_2{\rm Se}_3$
has a much smaller band gap than hBN, and therefore much larger values
of the static in-plane and out-of-plane permittivities. The fact that
$v^*_{\rm F}|_{\text{hyperbolic}}$ is larger than $v^*_{\rm
F}|_{\text{non-hyperbolic}}$ is due to the fact that ${\rm Bi}_2{\rm
Se}_3$ has reststrahlen bands at much lower energies,
the lower one
occurring at energies on the order of $8~{\rm meV}$.\footnote{%
At these low energies, one may worry about thermal occupation of the phonon-polariton modes. Since $8~{\rm meV}$ corresponds to a temperature on the order of $100~{\rm K}$, in order to measure the quasiparticle velocity $v_{\text{F}}^*$ of
graphene without worrying about these thermal effects,
SdH oscillations need to be measured up to a maximum temperature
$T_{\text{max}} \sim \SI{100}{K}$. In order to extract $v^*_{\text{F}}$ from a fit of the $T$-dependence of the amplitude of the SdH oscillations, one also needs to make sure that in the $0~{\rm K} < T < 100~{\rm K}$ range, $A(B,T)$ changes appreciably. 
From Eq.~(\ref{eq:amplitude}), we see that $A(B,T)$ decreases by a factor of $10$
when the argument of the hyperbolic sinus is
$2\pi^2k_{\text{B}}T/\hbar\omega_{\text{c}}\sim 5$.
Thus, the highest $T$ one should reach is estimated to be
$T_{\text{theory}}\sim(5eBv_{\text{F}})/
(2\pi^{5/2}k_{\text{B}}c\lvert n\rvert^{1/2})$.
In SI units,
$T_{\text{theory}}\sim(\bar B\lvert\bar n\rvert^{-1/2})\,\SI{20}{K}$,
where $\bar B$ is the applied magnetic field in Tesla and $\bar n$ the
carrier density in units of $\SI{e12}{cm^{-2}}$.
For typical densities and applied magnetic fields~\cite{Novoselov_Nature_2005,Elias_NatPhys_2011}, this theoretical estimate is also on the order of $\SI{100}{K}$. We can therefore neglect thermal occupation of phonon-polariton modes in the case of ${\rm Bi}_2{\rm Se}_3$.
}
Fig.~\ref{fig:percentage} shows the quantity $\Delta_{\rm QED}$ as a
function of system parameters. Without fine tuning, the QED effect on
the quasiparticle Fermi velocity is much larger in this cavity rather
than in the case of hBN.

A systematic quest of hyperbolic dielectrics with reststrahlen bands at ultra-low energies in order to maximize $\Delta_{\rm QED}$ is beyond the scope of the present Article. Nevertheless, our work shows that measurable QED effects emerge in DC magneto-transport in planar van der Waals cavities. We expect that these effects can be greatly enhanced in cavities where SHPPs and matter degrees of freedom are subject to lateral (and not only vertical) confinement, as in the case of Ref.~\cite{Hanan_arXiv_2022}. 
Recently developed on-chip THz spectroscopy~\cite{Kipp_arxiv_2024} may also be used to reveal strong coupling effects in van de Waals heterostructures.
From a theoretical point of view, it will be very interesting to extend the theory presented in this work to (a) real and moir\'{e} crystals and (b) strongly correlated materials~\cite{Loon_npj_2023}.

\paragraph*{Data, Materials, and Software Availability.}
The data that support the ﬁndings of this article are openly
available~\cite{code}.

\paragraph*{Acknowledgements.}
M. P. wishes to thank G. M. Andolina, M. Ceccanti, F. H. L. Koppens, and
A. R. Hamilton for many useful discussions.
R. R. and A. T. acknowledge the ``National Centre for HPC, Big Data and
Quantum Computing'' under the National Recovery and Resilience Plan
(NRRP), Mission 4 Component 2 Investment 1.4 CUP I53C22000690001 funded
from the European Union -- NextGenerationEU.
G. M. was supported by the Rita-Levi Montalcini program of the MUR.  A. H. M. was supported by a grant from the Keck foundation. M. P. was supported by the European Union's Horizon 2020 research and innovation programme under the Marie Sklodowska-Curie grant agreement No.~873028 - Hydrotronics and by the MUR - Italian Minister of University and Research under the ``Research projects of relevant national interest  - PRIN 2020''  - Project No.~2020JLZ52N, title ``Light-matter interactions and the collective behavior of quantum 2D materials (q-LIMA)''.


\begin{thebibliography}{199}
%
\bibitem{Shoenberg}
D. Shoenberg, \href{https://doi.org/10.1017/CBO9780511897870}{\textit{Magnetic Oscillations in Metals}} (Cambridge
University Press, Cambridge, 1984).
%
\bibitem{Landau09}
L.~D. Landau and E.~M. Lifshitz, \textit{Course of Theoretical Physics: Statistical Physics, Part 2} (Pergamon, New York, 1980).
%
\bibitem{Ando_RMP_1982}
T. Ando, A.~B. Fowler, and F. Stern, Electronic properties of two-dimensional systems, 
\href{https://doi.org/10.1103/RevModPhys.54.437}{Rev. Mod. Phys.~\textbf{54}, 437 (1982)}.
%
\bibitem{Rice_AnnPhys_1965}
T.~M Rice, The effects of electron-electron interaction on the properties of metals, 
\href{https://doi.org/10.1016/0003-4916(65)90234-4}{Ann. Phys. (N.Y.)~\textbf{31}, 100 (1965)}.
%
\bibitem{Pines_and_Nozieres} 
D. Pines and P. Nozi\'eres, \href{https://doi.org/10.4324/9780429492662}{\textit{The Theory of Quantum Liquids}} (W.A. Benjamin, Inc., New York, 1966).
%
\bibitem{Mahan_2000} 
G.~D. Mahan, \href{https://doi.org/10.1007/978-1-4757-5714-9}{\textit{Many-Particle Physics, Third Edition}} (Plenum Publishers, New York, 2000).
%
\bibitem{Giuliani_and_Vignale}
G.~F. Giuliani and G. Vignale, \href{https://doi.org/10.1017/CBO9780511619915}{\textit{Quantum Theory of the Electron Liquid}} (Cambridge University Press, Cambridge, 2005).
%
\bibitem{Fang_PhysRev_1968}
F.~F. Fang and P.~J. Stiles, Effects of a tilted magnetic field on a two-dimensional electron gas, 
\href{https://doi.org/10.1103/PhysRev.174.823}{Phys. Rev.~\textbf{174}, 823 (1968)}.
%
\bibitem{Smith_PRL_1972}
J.~L. Smith and P.~J. Stiles, Electron-electron interactions continuously variable in the range $2.1 > r_{\rm s} > 0.9$, 
\href{https://doi.org/10.1103/PhysRevLett.29.102}{Phys. Rev. Lett.~\textbf{29}, 102 (1972)}.
%
\bibitem{Pudalov_PRL_2002}
V.~M. Pudalov, M.~E. Gershenson, H. Kojima, N. Butch, E.~M. Dizhur, G. Brunthaler, A. Prinz, and G. Bauer, Low-density spin susceptibility and effective mass of mobile electrons in Si inversion layers, \href{https://doi.org/10.1103/PhysRevLett.88.196404}{Phys. Rev. Lett.~\textbf{88}, 196404 (2002)}.
%
\bibitem{Vakili_PRL_2004}
K. Vakili, Y.~P. Shkolnikov, E. Tutuc, E.~P. De Poortere, and M. Shayegan, Spin susceptibility of two-dimensional electrons in narrow AlAs quantum wells, \href{https://doi.org/10.1103/PhysRevLett.92.226401}{Phys. Rev. Lett.~\textbf{92}, 226401 (2004)}.
%
\bibitem{Padmanabhan_PRL_2008}
M. Padmanabhan, T. Gokmen, N.~C. Bishop, and M. Shayegan, 
Effective mass suppression in dilute, spin-polarized two-dimensional electron systems,
\href{https://doi.org/10.1103/PhysRevLett.101.026402}{Phys. Rev. Lett.~\textbf{101}, 026402 (2008)}.
%
\bibitem{Gokmen_PRL_2008}
T. Gokmen, M. Padmanabhan, and M. Shayegan, Dependence of effective mass on spin and valley degrees of freedom, 
\href{https://doi.org/10.1103/PhysRevLett.101.146405}{Phys. Rev. Lett.~\textbf{101}, 146405 (2008)}.
%
\bibitem{Gokmen_PRB_2009}
T. Gokmen, M. Padmanabhan, K. Vakili, E. Tutuc, and M. Shayegan, Effective mass suppression upon complete spin-polarization in an isotropic two-dimensional electron system, \href{https://doi.org/10.1103/PhysRevB.79.195311}{Phys. Rev. B~\textbf{79}, 195311 (2009)}.
%
\bibitem{Tan_PRL_2005}
Y.-W. Tan, J. Zhu, H.~L. Stormer, L.~N. Pfeiffer, K.~W. Baldwin, and K.~W. West, Measurements of the density-dependent many-body electron mass in two dimensional GaAs/AlGaAs heterostructures, \href{https://doi.org/10.1103/PhysRevLett.94.016405}{Phys. Rev. Lett.~\textbf{94}, 016405 (2005)}.
%
\bibitem{Tan_PhysicaE_2006}
Y.-W. Tan, J. Zhu, H.~L. Stormer, L.~N. Pfeiffer, K.~W. Baldwin, and K.~W. West, 
Spin susceptibility and effective mass of a 2D electron system in GaAs heterostructures towards the weak interacting regime, 
\href{https://doi.org/10.1016/j.physe.2006.03.070}{Physica E: Low Dimens. Syst. Nanostruct.~\textbf{34} 260 (2006)}.
%
\bibitem{Chiu_PRB_2011}
YenTing Chiu, M. Padmanabhan, T. Gokmen, J. Shabani, E. Tutuc, M. Shayegan, and R. Winkler, Effective mass and spin susceptibility of dilute two-dimensional holes in GaAs, \href{https://doi.org/10.1103/PhysRevB.84.155459}{Phys. Rev. B~\textbf{84}, 155459 (2011)}.
%
\bibitem{Caviglia_PRL_2010}
A.~D. Caviglia, S. Gariglio, C. Cancellieri, B. Sac\'{e}p\'{e}, A. F\^{e}te, N. Reyren, M. Gabay, A.~F. Morpurgo, and J.-M. Triscone, 
Two-dimensional quantum oscillations of the conductance at ${\rm LaAlO}_3/{\rm SrTiO}_3$ interfaces,
\href{https://doi.org/10.1103/PhysRevLett.105.236802}{Phys. Rev. Lett.~\textbf{105}, 236802 (2010)}.
%
\bibitem{BenShalom_PRL_2010}
M. Ben Shalom, A. Ron, A. Palevski, and Y. Dagan, Shubnikov-de Haas oscillations in ${\rm LaAlO}_3/{\rm SrTiO}_3$ interface,
\href{https://doi.org/10.1103/PhysRevLett.105.206401}{Phys. Rev. Lett.~\textbf{105}, 206401 (2010)}.
%
\bibitem{Melnikov_SciRep_2023}
M.~Yu. Melnikov, A.~A. Shakirov, A.~A. Shashkin, S.~H. Huang, C.~W. Liu, and S.~V. Kravchenko, Spin independence of the strongly enhanced effective mass in ultra-clean SiGe/Si/SiGe two-dimensional electron system, \href{https://doi.org/10.1038/s41598-023-44580-y}{Sci. Rep.~\textbf{13}, 17364 (2023)}.
%
\bibitem{Novoselov_Nature_2005}
K.~S. Novoselov, A.~K. Geim, S.~V. Morozov, D. Jiang, M.~I. Katsnelson, I.~V. Grigorieva, S.~V. Dubonos, and A.~A. Firsov, Two-dimensional gas of massless Dirac fermions in graphene, \href{https://doi.org/10.1038/nature04233}{Nature~\textbf{438}, 197 (2005)}.
%
\bibitem{Zhang_Nature_2005}
Y. Zhang, Y.-W. Tan, H.~L. Stormer, and P. Kim, Experimental observation of the quantum Hall effect and Berry's phase in graphene, \href{https://doi.org/10.1038/nature04235}{Nature~\textbf{438}, 201 (2005)}.
%
\bibitem{Elias_NatPhys_2011}%SdH SLG
D.~C. Elias, R.~V. Gorbachev, A.~S. Mayorov, S.~V. Morozov, A.~A. Zhukov, P. Blake, L.~A. Ponomarenko, I.~V. Grigorieva, K.~S. Novoselov, F. Guinea, and A.~K. Geim, Dirac cones reshaped by interaction effects in suspended graphene. 
\href{https://doi.org/10.1038/nphys2049}{Nat.~Phys.~\textbf{7}, 701 (2011)}.
%	
\bibitem{Zou_PRB_2011}%SdH BLG
K. Zou, X. Hong, and J. Zhu, Effective mass of electrons and holes in bilayer graphene: Electron-hole asymmetry and electron-electron interaction, \href{https://doi.org/10.1103/PhysRevB.84.085408}{Phys. Rev. B~\textbf{84}, 085408 (2011)}.
%
\bibitem{Fallahazad_PRL_2016}
B. Fallahazad, H.~C.~P. Movva, K. Kim, S. Larentis, T. Taniguchi, K. Watanabe, S.~K. Banerjee, and E. Tutuc, 
Shubnikov-de Haas oscillations of high-mobility holes in monolayer and bilayer ${\rm WSe}_2$: Landau level degeneracy, effective mass, and negative compressibility, \href{https://doi.org/10.1103/PhysRevLett.116.086601}{Phys. Rev. Lett.~\textbf{116}, 086601 (2016)}.
%
\bibitem{Novotny_2012}
L. Novotny and B. Hecht, \href{https://doi.org/10.1017/CBO9780511794193}{\textit{Principles of Nano-Optics, Second Edition}} (Cambridge University Press, Cambridge, 2012). 
%
\bibitem{Basov_Nanophotonics_2021}
D.~N. Basov, A. Asenjo-Garcia, P.~J. Schuck, X. Zhu, and A. Rubio, Polariton panorama, \href{https://doi.org/10.1515/nanoph-2020-0449}{Nanophoton.~\textbf{10}, 549 (2021)}.
%
\bibitem{Plantey_ACS_2021} 
A. Reserbat-Plantey, I. Epstein, I. Torre, A.~T. Costa, P.~A.~D. Gon\c{c}alves, N.~A. Mortensen, M. Polini, J.~C.~W. Song, N.~M.~R. Peres, and F.~H.~L. Koppens, Quantum nanophotonics in two-dimensional materials, \href{https://doi.org/10.1021/acsphotonics.0c01224}{ACS Photon.~\textbf{8}, 85 (2021)}.
%
\bibitem{Bloch_Nature_2022}
J. Bloch, A. Cavalleri, V. Galitski, M. Hafezi, and A. Rubio, Strongly correlated electron–photon systems, \href{https://doi.org/10.1038/s41586-022-04726-w}{Nature~\textbf{606}, 41 (2022)}.
%
\bibitem{Schlawin_APR_2022}
F. Schlawin, D.~M. Kennes, and M.~A. Sentef, Cavity quantum materials, \href{https://doi.org/10.1063/5.0083825}{Appl. Phys. Rev.~\textbf{9}, 011312 (2022)}.
%
\bibitem{GarciaVidal_Science_2021}
F.~J. Garcia-Vidal, C. Ciuti, and T.~W. Ebbesen, Manipulating matter by strong coupling to vacuum fields, \href{https://doi.org/10.1126/science.abd0336}{Science~\textbf{373}, 6551 (2021)}.
%
\bibitem{Genet_PT_2021}
C. Genet, J. Faist, and T. Ebbesen, Inducing new material properties with hybrid light–matter states, \href{https://doi.org/10.1063/PT.3.4749}{Phys. Today~\textbf{74}, 42 (2021)}.
%
\bibitem{Rubio_NatureMater_2021}
H. H\"{u}bener, U. De Giovannini, C. Sch\"{a}fer, J. Andberger, M. Ruggenthaler, J. Faist, and A. Rubio, Engineering quantum materials with chiral optical cavities, \href{https://doi.org/10.1038/s41563-020-00801-7}{Nat. Mater.~\textbf{20}, 438 (2021)}.
%
\bibitem{Bartolo_PRB_2018}
N. Bartolo and C. Ciuti, Vacuum-dressed cavity magnetotransport of a two-dimensional electron gas, 
\href{https://doi.org/10.1103/PhysRevB.98.205301}{Phys. Rev. B~\textbf{98}, 205301 (2018)}.
%
\bibitem{Ciuti_PRB_2021}
C. Ciuti, Cavity-mediated electron hopping in disordered quantum Hall systems, 
\href{https://doi.org/10.1103/PhysRevB.104.155307}{Phys. Rev. B~\textbf{104}, 155307 (2021)}.
%
\bibitem{Arwas_PRB_2023}
G. Arwas and C. Ciuti, Quantum electron transport controlled by cavity vacuum fields, 
\href{https://doi.org/10.1103/PhysRevB.107.045425}{Phys. Rev. B~\textbf{107}, 045425 (2023)}.
%
\bibitem{scalari_science_2012}
G. Scalari, C. Maissen, D. Turcinková, D. Hagenmüller, S. De Liberato, C. Ciuti, C. Reichl, D. Schuh, W. Wegscheider, 
M. Beck, and J. Faist, Ultrastrong coupling of the cyclotron transition of a 2D electron gas to a THz metamaterial, \href{https://doi.org/10.1126/science.1216022}{Science~\textbf{335}, 1323 (2012)}.
%
\bibitem{muravev_prb_2013}
V.~M. Muravev, P.~A. Gusikhin, I.~V. Andreev, and I.~V. Kukushkin, Ultrastrong coupling of high-frequency two-dimensional cyclotron plasma mode with a cavity photon, \href{https://doi.org/10.1103/PhysRevB.87.045307}{Phys. Rev. B~\textbf{87}, 045307 (2013)}.
%
\bibitem{maissen_prb_2014} 
C. Maissen, G. Scalari, F. Valmorra, M. Beck, J. Faist, S. Cibella, R. Leoni, C. Reichl, C. Charpentier, and W. Wegscheider, Ultrastrong coupling in the near field of complementary split-ring resonators, \href{https://doi.org/10.1103/PhysRevB.90.205309}{Phys. Rev. B \textbf{90}, 205309 (2014)}.
%
\bibitem{smolka_science_2014}
S. Smolka, W. Wuester, F. Haupt, S. Faelt, W. Wegscheider, and A. Imamoglu, Cavity quantum electrodynamics with many-body states of a two-dimensional electron gas, \href{https://doi.org/10.1126/science.1258595}{Science~\textbf{346}, 332 (2014)}.
%
\bibitem{Keller_nanolett_2017}
J. Keller, G. Scalari, S. Cibella, C. Maissen, F. Appugliese, E. Giovine, R. Leoni, M. Beck, and J. Faist, Few-electron ultrastrong light-matter coupling at $300~{\rm GHz}$ with nanogap hybrid LC microcavities, \href{https://doi.org/10.1021/acs.nanolett.7b03228}{Nano Lett.~\textbf{17}, 7410 (2017)}.
%
\bibitem{Paravicini_Bagliani_NaturePhys_2019}
G.~L. Paravicini-Bagliani, F. Appugliese, E. Richter, F. Valmorra, J. Keller, M. Beck, N. Bartolo, C. R\"{o}ssler, T. Ihn, K. Ensslin, C. Ciuti, G. Scalari, and J. Faist, Magneto-transport controlled by Landau polariton states, \href{https://doi.org/10.1038/s41567-018-0346-y}{Nat. Phys.~\textbf{15}, 186 (2019)}.
%
\bibitem{Appugliese_science_2022}
F. Appugliese, J. Enkner, G.~L. Paravicini-Bagliani, M. Beck, C. Reichl, W. Wegscheider, G. Scalari, C. Ciuti, and J. Faist, Breakdown of topological protection by cavity vacuum fields in the integer quantum Hall effect, \href{https://doi.org/10.1126/science.abl5818}{Science~\textbf{375}, 6584 (2022)}.
\bibitem{Rokaj_2023}
V. Rokaj, J. Wang, J. Sous, M. Penz, M. Ruggenthaler, and A. Rubio,
Weakened Topological Protection of the Quantum Hall Effect in a Cavity,
\href{https://doi.org/10.1103/PhysRevLett.131.196602}{Phys. Rev. Lett. \textbf{131}, 196602 (2023)}.
%
%
\bibitem{Lifshitz_1955}
I.~M. Lifshitz and A.~M. Kosevich, Theory of Magnetic Susceptibility in Metals at Low Temperatures, Zh. Eksp. Teor. Fiz.~\textbf{29}, 730 (1955) [\href{http://jetp.ras.ru/cgi-bin/dn/e_002_04_0636.pdf}{Sov. Phys. JETP~\textbf{2}, 636 (1956)}].
%
\bibitem{Isihara_JPhysC_1986}
A. Isihara and L. Smr\v{c}ka, Density and magnetic field dependences of the conductivity of two-dimensional electron systems, \href{https://doi.org/10.1088/0022-3719/19/34/015}{J. Phys. C~\textbf{19}, 6777 (1986)}.
%
\bibitem{Sharapov_PRB_2004}
S.~G. Sharapov, V.~P. Gusynin, and H. Beck, Magnetic oscillations in planar systems with the Dirac-like spectrum of quasiparticle excitations, \href{https://doi.org/10.1103/PhysRevB.69.075104}{Phys. Rev. B~\textbf{69}, 075104 (2004)}.
%
\bibitem{Kopelevich_PRL_2004}
I.~A. Luk'yanchuk and Y. Kopelevich, Phase analysis of quantum oscillations in graphite, 
\href{https://doi.org/10.1103/PhysRevLett.93.166402}{Phys. Rev. Lett.~\textbf{93}, 166402 (2004)}.
%
\bibitem{Kuppersbusch_PRB_2017}
C. K\"{u}ppersbusch and L. Fritz, Modifications of the Lifshitz-Kosevich formula in two-dimensional Dirac systems, \href{https://doi.org/10.1103/PhysRevB.96.205410}{Phys. Rev. B~\textbf{96}, 205410 (2017)}.
%
\bibitem{Mikitik_PRL_1999}
G.~P. Mikitik and Yu.~V. Sharlai, Manifestation of Berry's phase in metal physics, \href{https://doi.org/10.1103/PhysRevLett.82.2147}{Phys. Rev. Lett.~\textbf{82}, 2147 (1999)}.
%
\bibitem{Sun_ACS_2014}
J. Sun, N.~M. Litchinitser, and J. Zhou, 
Indefinite by nature: From ultraviolet to terahertz,  
\href{https://doi.org/10.1021/ph4000983}{ACS Photon.~\textbf{1}, 293 (2014)}.
%
\bibitem{Basov_Science_2016}
D.~N. Basov, M.~M. Fogler, and F.~J.~G. de Abajo, Polaritons in van der Waals materials, 
\href{https://doi.org/10.1126/science.aag1992}{Science~\textbf{354}, aag1992 (2016)}.
%
\bibitem{Low_NatMater_2017}
T. Low, A. Chaves, J.~D. Caldwell, A. Kumar, N.~X. Fang, P. Avouris, T.~F. Heinz, F. Guinea, L. Martin-Moreno, and 
F. Koppens, Polaritons in layered two-dimensional materials, 
\href{https://doi.org/10.1038/nmat4792}{Nat. Mater.~\textbf{16}, 182 (2017)}.
%
\bibitem{Zhang_Nature_2021}
Q. Zhang, G. Hu, W. Ma, P. Li, A. Krasnok, R. Hillenbrand, A. Al\`u, and 
C.-W. Qiu, Interface nano-optics with van der Waals polaritons, 
\href{https://doi.org/10.1038/s41586-021-03581-5}{Nature~\textbf{597}, 187 (2021)}.
%
\bibitem{Dai_science_2014}
S. Dai, Z. Fei, Q. Ma, A.~S. Rodin, M. Wagner, A.~S. McLeod, M.~K. Liu, W. Gannett, W. Regan, K. Watanabe, T. Taniguchi, M. Thiemens, G. Dominguez, A.~H. Castro Neto, A. Zettl, F. Keilmann, P. Jarillo-Herrero, M.M. Fogler, and D.~N. Basov, Tunable phonon polaritons in atomically thin van der Waals crystals of boron nitride, \href{http://dx.doi.org/10.1126/science.1246833}{Science~\textbf{343}, 1125 (2014)}.
%
\bibitem{Caldwell_naturecommun_2014}
J.~D. Caldwell, A. Kretinin, Y. Chen, V. Giannini, M.~M. Fogler, Y. Francescato, C.~T. Ellis, J.~G. Tischler, C.~R. Woods, A.~J. Giles, M. Hong, K. Watanabe, T. Taniguchi, S.~A. Maier, and K.~S. Novoselov, Sub-diffractional volume-confined polaritons in the natural hyperbolic material hexagonal boron nitride, \href{http://dx.doi.org/10.1038/ncomms6221}{Nat. Commun.~\textbf{5}, 5221 (2014)}.
%
\bibitem{Li_NatCommun_2015}
P. Li, M. Lewin, A.~V. Kretinin, J.~D. Caldwell, K.~S. Novoselov, T. Taniguchi, K. Watanabe, F. Gaussmann, and T. Taubner, Hyperbolic phonon-polaritons in boron nitride for near-field optical imaging and focusing, \href{https://doi.org/10.1038/ncomms8507}{Nat. Commun.~\textbf{6}, 7507 (2015)}.
%
\bibitem{Dai_NatCommun_2015}
S. Dai, Q. Ma, T. Andersen, A.~S. McLeod, Z. Fei, M.~K. Liu, M. Wagner, K. Watanabe, T. Taniguchi, M. Thiemens, F. Keilmann, P. Jarillo-Herrero, M.~M. Fogler, and D.~N. Basov, Subdiffractional focusing and guiding of polaritonic rays in a natural hyperbolic material, \href{https://doi.org/10.1038/ncomms7963}{Nat. Commun.~\textbf{6}, 6963 (2015)}.
%
\bibitem{Eva_NatCommun_2021}
E.~A.~A. Pogna, L. Viti, A. Politano, M. Brambilla, G. Scamarcio, and M.~S. Vitiello, 
Mapping propagation of collective modes in ${\rm Bi}_2{\rm Se}_3$ and ${\rm Bi}_2{\rm Te}_{2.2}{\rm Se}_{0.8}$ 
topological insulators by near-field terahertz nanoscopy, \href{https://doi.org/10.1038/s41467-021-26831-6}{Nat. Commun.~\textbf{12}, 6672 (2021)}.
%
\bibitem{Cohen_1997}
C. Cohen-Tannoudji, J. Dupont-Roc, and G. Grynberg,
\href{https://doi.org/10.1002/9783527618422}{\textit{Photons and Atoms: Introduction to Quantum Electrodynamics}}
(Wiley-VCH, Berlin, 1997).
%
\bibitem{andolina_arxiv_2022}
G.~M. Andolina, A. De Pasquale, F.~M.~D. Pellegrino, I. Torre, F.~H.~L. Koppens, and M. Polini, Amperean superconductivity cannot be induced by deep subwavelength cavities in a two-dimensional material, \href{https://doi.org/10.1103/PhysRevB.109.104513}{Phys. Rev. B~\textbf{109}, 104513 (2024)}.
%
\bibitem{Flick_ACS_2019}
J. Flick, D.~M. Welakuh, M. Ruggenthaler, H. Appel, and A. Rubio, 
Light-matter response in nonrelativistic quantum electrodynamics, 
\href{https://doi.org/10.1021/acsphotonics.9b00768}{ACS Photon.~\textbf{6}, 2757 (2019)}.
%
\bibitem{Amelio_PRB_2021}
I. Amelio, L. Korosec, I. Carusotto, and G. Mazza, Optical dressing of the electronic response of two-dimensional semiconductors in quantum and classical descriptions of cavity electrodynamics, \href{https://doi.org/10.1103/PhysRevB.104.235120}{Phys. Rev. B~\textbf{104}, 235120 (2021)}.
%
\bibitem{nataf_prl_2019}
P.  Nataf, T. Champel, G. Blatter, and D.~M. Basko, Rashba cavity QED: A route towards the superradiant quantum phase transition, \href{https://doi.org/10.1103/PhysRevLett.123.207402}{Phys. Rev. Lett.~\textbf{123}, 207402 (2019)}.
%
\bibitem{andolina_prb_2020}
G.~M. Andolina, F.~M.~D. Pellegrino, V. Giovannetti, A.~H. MacDonald, and M. Polini, Theory of photon condensation in a spatially varying electromagnetic field, \href{https://doi.org/10.1103/PhysRevB.102.125137}{Phys. Rev. B~\textbf{102}, 125137 (2020)}.
%
\bibitem{guerci_prl_2020}
D. Guerci, P. Simon, and C. Mora, Superradiant phase transition in electronic systems and emergent topological phases, \href{https://doi.org/10.1103/PhysRevLett.125.257604}{Phys. Rev. Lett.~\textbf{125}, 257604 (2020)}.
%
\bibitem{mazza_prb_2023}
G. Mazza and M. Polini, Hidden excitonic quantum states with broken time-reversal symmetry, 
\href{https://doi.org/10.1103/PhysRevB.108.L241107}{Phys. Rev. B~\textbf{108}, L241107 (2023)}.
%
\bibitem{Asenjo-Garcia_2017}
A. Asenjo-Garcia, J.~D. Hood, D.~E. Chang, and H.~J. Kimble,
Atom-light interactions in quasi-one-dimensional nanostructures: A Green's-function perspective,
\href{https://doi.org/10.1103/PhysRevA.95.033818}{%
Phys. Rev. A \textbf{95}, 033818 (2017).}
%
\bibitem{footnote_spatial_nonlocality}
L.~D. Landau and E.~M. Lifshitz, \textit{Course of Theoretical Physics: Electrodynamics of Continuos Media} (Pergamon, New York, 1984).
%
\bibitem{Jackson}
J.~D. Jackson, \textit{Classical Electrodynamics, Third Edition} (Wiley, New York, 1999).
%
\bibitem{Glauber_PRA_1991}
R.~J. Glauber and M. Lewenstein, Quantum optics of dielectric media, \href{https://doi.org/10.1103/PhysRevA.43.467}{Phys. Rev. A~\textbf{43}, 467 (1991)}.
%
\bibitem{Hedin_1969}
L. Hedin and S. Lundqvist, Effects of electron-electron and electron-phonon interactions on the one-electron states of solids, \href{https://doi.org/10.1016/S0081-1947(08)60615-3}{Solid State Phys.~\textbf{23}, 1 (1969)}.
%
\bibitem{Geim_Nature_2013}
A. Geim and I. Grigorieva, Van der Waals heterostructures, \href{https://doi.org/10.1038/nature12385}{Nature~\textbf{499}, 419 (2013)}.
%
\bibitem{Keldysh_1979}
L.~V. Keldysh, Coulomb interaction in thin semiconductor and semimetal films, 
\href{http://jetpletters.ru/ps/1458/article_22207.pdf}{Pis'ma Zh. Eksp. Teor. Fiz.~\textbf{29}, 716 (1979)}.
%
\bibitem{Tomadin_prl_2015}
A. Tomadin, A. Principi, J.~C.~W. Song, L.~S. Levitov, and M. Polini,
Accessing phonon polaritons in hyperbolic crystals by angle-resolved photoemission spectroscopy, \href{https://doi.org/10.1103/PhysRevLett.115.087401}{Phys. Rev. Lett.~\textbf{115}, 087401 (2015)}.
%
\bibitem{Woessner_naturemater_2015}
A. Woessner, M.~B. Lundeberg, Y. Gao, A. Principi, P. Alonso-Gonz\'alez, M. Carrega, K. Watanabe, T. Taniguchi, G. Vignale, M. Polini, J. Hone, R. Hillenbrand, and F.~H.~L. Koppens, Highly confined low-loss plasmons in graphene–boron nitride heterostructures, \href{http://dx.doi.org/10.1038/nmat4169}{Nat. Mater.~\textbf{14}, 421 (2015)}.
%
\bibitem{Lundeberg_Science_2017}
M.~B. Lundeberg, Y. Gao, R. Asgari, C. Tan, B. Van Duppen, M. Autore, P. Alonso-Gonz\'{a}lez, 
A. Woessner, K. Watanabe, T. Taniguchi, R. Hillenbrand, J. Hone, M. Polini, and F.~H.~L. Koppens, Tuning quantum nonlocal effects in graphene plasmonics, 
\href{http://dx.doi.org/10.1126/science.aan2735}{Science~\textbf{357}, 187 (2017)}.
%
\bibitem{Alonso_NatureNano_2017}
P. Alonso-Gonz\'{a}lez, A.~Y. Nikitin, Y. Gao, A. Woessner, M.~B. Lundeberg, A. Principi, N. Forcellini, W. Yan, S. V\'{e}lez, A.~J. Huber, K. Watanabe, T. Taniguchi, F. Casanova, L.~E. Hueso, M. Polini, J. Hone, F.~H.~L. Koppens, and R. Hillenbrand, Ultra-confined acoustic THz graphene plasmons revealed by photocurrent nanoscopy, \href{http://dx.doi.org/10.1038/nnano.2016.185}{Nat. Nanotech.~\textbf{12}, 31 (2017)}.
%
\bibitem{Kim_NatureCommun_2020}
M. Kim, S.~G. Xu, A.~I. Berdyugin, A. Principi, S. Slizovskiy, N. Xin, P. Kumaravadivel, W. Kuang, M. Hamer, R. Krishna Kumar, R.~V. Gorbachev, K. Watanabe, T. Taniguchi, I.~V. Grigorieva, V.~I. Fal'ko, M. Polini, and A.~K. Geim, Control of electron-electron interaction in graphene by proximity screening, \href{https://doi.org/10.1038/s41467-020-15829-1}{Nat. Commun.~\textbf{11}, 2339 (2020)}.
%
\bibitem{Kotov_RMP_2012} 
V.~N. Kotov, B. Uchoa, V.~M. Pereira, F. Guinea, and A.~H. Castro Neto, Electron-electron interactions in graphene: current status and perspectives, \href{https://doi.org/10.1103/RevModPhys.84.1067}{Rev. Mod. Phys.~\textbf{84}, 1067 (2012)}.
%
\bibitem{Quinn_Ferrell}
J.~J. Quinn and R.~A. Ferrell, Electron self-energy approach to correlation in a degenerate electron gas, \href{https://doi.org/10.1103/PhysRev.112.812}{Phys. Rev.~\textbf{112}, 812 (1958)}.
%
\bibitem{Hedin_physrev_1965}
L. Hedin, New method for calculating the one-particle Green's function with application to the electron-gas problem, 
\href{https://doi.org/10.1103/PhysRev.139.A796}{Phys. Rev.~\textbf{139}, A796 (1965)}.
%
\bibitem{Hybertsen_PRB_1986}
M.~S. Hybertsen and S.~G. Louie, 
Electron correlation in semiconductors and insulators: Band gaps and quasiparticle energies, 
\href{https://doi.org/10.1103/PhysRevB.34.5390}{Phys. Rev. B~\textbf{34}, 5390 (1986)}.
%
\bibitem{G0W-validity}
B. Holm and U. von Barth,
Fully self-consistent GW self-energy of the electron gas,
\href{https://doi.org/10.1103/PhysRevB.57.2108}{Phys. Rev. B~\textbf{57},
2108 (1998)}.

%
\bibitem{Polini_ssc_2007}
M. Polini, R. Asgari, Y. Barlas, T. Pereg-Barnea, and A.~H. MacDonald, Graphene: a pseudo-chiral Fermi liquid, \href{http://dx.doi.org/10.1016/j.ssc.2007.04.035}{Solid State Commun.~\textbf{143}, 58 (2007)}.
%
\bibitem{Polini_prb_2008}
M. Polini, R. Asgari, G. Borghi, Y. Barlas, T. Pereg-Barnea, and A.~H. MacDonald, Plasmons and the spectral function of graphene, 
\href{http://dx.doi.org/10.1103/PhysRevB.77.081411}{Phys. Rev. B~\textbf{77}, 081411(R) (2008)}. 
%
\bibitem{Barlas_PRL_2007}
Y. Barlas, T. Pereg-Barnea, M. Polini, R. Asgari, and A.~H. MacDonald, Chirality and correlations in graphene, \href{https://doi.org/10.1103/PhysRevLett.98.236601}{Phys. Rev. Lett.~\textbf{98}, 236601 (2007)}.
%
\bibitem{Bostwick_2010}
A. Bostwick, F. Speck, T. Seyller, K. Horn, M. Polini, R. Asgari,
A.~H. MacDonald, and E. Rotenberg,
Observation of Plasmarons in Quasi-Freestanding Doped Graphene,
\href{http://dx.doi.org/10.1126/science.1186489}{%
Science \textbf{328}, 999 (2010)}.

\bibitem{Walter_2011}
A.~L. Walter, A. Bostwick, K. Jeon, F. Speck, M. Ostler, T. Seyller,
L. Moreschini, Y.~J. Chang, M. Polini, R. Asgari, A.~H. MacDonald,
K. Horn, and E. Rotenberg,
Effective screening and the plasmaron bands in graphene,
\href{http://dx.doi.org/10.1103/PhysRevB.84.085410}{%
Phys. Rev. B \textbf{84}, 085410 (2011)}.

%
\bibitem{Shung_PRB_1986}
K.~W.-K.~Shung, Dielectric function and plasmon structure of stage-1 intercalated graphite, \href{https://doi.org/10.1103/PhysRevB.34.979}{Phys. Rev. B~\textbf{34}, 979 (1986)}.
%
\bibitem{Wunsch_NJP_2016}
B. Wunsch, T. Stauber, F. Sols, and F. Guinea, Dynamical polarization of graphene at finite doping, 
\href{https://doi.org/10.1088/1367-2630/8/12/318}{New J. Phys.~\textbf{8}, 318 (2006)}.
%
\bibitem{Hwang_PRB_2006}
E.~H. Hwang and S. Das Sarma, Dielectric function, screening, and plasmons in two-dimensional graphene, \href{https://doi.org/10.1103/PhysRevB.75.205418}{Phys. Rev. B~\textbf{75}, 205418 (2007)}.
%
\bibitem{Ashida_PRL_2023}
Y. Ashida, A. $\dot{\text{I}}$mamo\u{g}lu, and E. Demler, Cavity quantum electrodynamics with hyperbolic van der Waals materials,
\href{https://doi.org/10.1103/PhysRevLett.130.216901}{Phys. Rev. Lett.~\textbf{130}, 216901 (2023)}.
%
%
\bibitem{Hanan_arXiv_2022}
H.~H. Sheinfux, L. Orsini, M. Jung, I. Torre, M. Ceccanti, S. Marconi,
R. Maniyara, D. Barcons Ruiz, A. H\"{o}tger, R. Bertini, S. Castilla,
N.~C.~H. Hesp, E. Janzen, A. Holleitner, V. Pruneri, J.~H. Edgar, G.
Shvets, and F.~H.~L. Koppens, High quality nanocavities through
multimodal confinement of hyperbolic polaritons in hexagonal boron
nitride, \href{https://doi.org/10.1038/s41563-023-01785-w}{Nat. Mater.
\textbf{23}, 499 (2024)}.
%
\bibitem{Kipp_arxiv_2024}
G. Kipp, H.~M. Bretscher, B. Schulte, D. Herrmann, K. Kusyak, M.~W. Day,
S. Kesavan, T. Matsuyama, X. Li, S.~M. Langner, J. Hagelstein, F. Sturm,
A.~M. Potts, C.~J. Eckhardt, Y. Huang, K. Watanabe, T. Taniguchi,
A. Rubio, D.~M. Kennes, M.~A. Sentef, E. Baudin, G. Meier,
M.~H. Michael, and J.~W. McIver,
Cavity electrodynamics of van der Waals heterostructures,
\href{https://doi.org/10.48550/arXiv.2403.19745}{arXiv:2403.19745}.
\bibitem{Loon_npj_2023}
E.~G.~C.~P. van Loon, M. Sch\"{u}ler, D. Springer, G. Sangiovanni, J.~M. Tomczak, and T.~O. Wehling, 
Coulomb engineering of two-dimensional Mott materials, \href{https://doi.org/10.1038/s41699-023-00408-x}{npj 2D Mater. Appl.~\textbf{7}, 47 (2023)} and references therein.

\bibitem{code}
Software used to produce the plots in this article is available at
Zenodo
\href{https://zenodo.org/records/15874135}{https://zenodo.org/records/15874135}.

\end{thebibliography}
\end{document}